\documentclass{emulateapj}
\usepackage{amssymb}
\shorttitle{Angular Power Spectrum of the EoR}
\shortauthors{Zheng et al.}

\begin{document}

\title{A Method to Extract the Angular Power Spectrum of 
the Epoch of Reionization from Low-Frequency Radio Interferometers}

\author{Qian Zheng\altaffilmark{1,2}, Xiang-Ping Wu\altaffilmark{1,3}, 
Jun-Hua Gu\altaffilmark{1}, Jingying Wang\altaffilmark{3}, 
and Haiguang Xu\altaffilmark{3}}

\altaffiltext{1}{National Astronomical Observatories, Chinese Academy of 
Sciences, Beijing 100012, China}
\altaffiltext{2}{Graduate School of Chinese Academy of Sciences, 
Beijing 100049, China}
\altaffiltext{3}{Department of Physics, Shanghai Jiao Tong University, 
800 Dongchuan Road, Shanghai 200240, China}

\begin{abstract}
The redshifted 21cm signal of neutral hydrogen from the epoch of 
reionization (EoR) is extremely weak and its first detection is 
therefore expected to be statistical with first-generation low-frequency 
radio interferometers. In this letter we propose a method to 
extract the angular power spectrum of EoR from the visibility 
correlation coefficients $p_{ij}(u,v)$, instead of the visibilities 
$V_{ij}(u,v)$ measured directly by radio interferometers in conventional
algorithm. The visibility correlation coefficients are defined as   
$p_{ij}(u,v)=V_{ij}(u,v)/\sqrt{|V_{ii}||V_{jj}|}$ by introducing 
the auto-correlation terms $V_{ii}$ and $V_{jj}$ such that the angular
power spectrum $C_{\ell}$ can be obtained through 
$C_{\ell}=4\pi^2 T_0^2\langle|p_{ij}(u,v)|^2\rangle$, independently of the 
primary beams of antennas. This also removes partially the influence of 
receiver gains in the measurement of $C_{\ell}$ because the
amplitudes of the gains cancel each other out in  
the statistical average operation of $\langle|p_{ij}(u,v)|^2\rangle$.
The average brightness temperature $T_0$ of extragalactic sources
is used as the calibrator of $C_{\ell}$. 
Finally we demonstrate the feasibility of the 
novel method using the simulated sky maps as targets and the 
21 CentiMeter Array (21CMA) as interferometer.  
\end{abstract}

\keywords{cosmology: theory --- diffuse radiation --- 
          intergalactic medium ---methods: data analysis--- techniques: interferometric }

\section{Introduction}

While the redshifted 21cm emission/absorption of 
neutral hydrogen provides a unique cosmological probe of the epoch of 
reionization (EoR) -, the last frontier of observational cosmology, 
there are two primary challenges of measuring 
the EoR signatures with most of the dedicated radio 
facilities (e.g., 21CMA, LOFAR, LWA, MWA, PAPER, SKA, etc.): First, 
the cosmic signal from EoR is deeply buried under the extremely bright 
foreground dominated by our Galaxy, extragalactic sources and 
telescope noise, and  an unprecedented level of foreground removals 
down to five orders of magnitude should be required in order to detect
the cosmic signal (e.g., Madau et al. 1997; 
Zaldarriaga et al. 2004). 
Second, low-frequency radio interferometric measurements in 
current 21cm experiments suffer from various instrumental 
contaminations in addition to man-made radio-frequency interference.
Outstanding among these are the frequency-dependent point 
spread function and field-of-view (also known as 'mode-mixing'),
complexity of calibration, and bright source subtraction (e.g., 
Morales et al. 2006; Liu et al. 2009;
Bowman et al. 2009; Datta et al. 2009,2010;
Bernardi et al. 2010; Datta et al. 2010; 
Petrovic \& Oh 2011). 
Yet,  it is generally agreed among the 21cm cosmology 
community that the advent of many sophisticated techniques and algorithms 
in recent years helps overcoming these observational and 
technical hurdles, allowing to reach the desired detection sensitivity
with the first generation of radio interferometers 
(for recent reviews see Furlanetto et al. 2006; Morales 
\& Wyithe 2010; Pritchard \& Loeb 2012; Zaroubi 2012).

The theoretically predicted brightness temperature of 21cm signal
from EoR is only $\sim10$ mK, and first capture of such extremely 
weak signal is therefore expected to be statistical (Zaldarriaga et al. 
2004). Most of the current 21cm 
experiments aiming to detect the EoR signal are based on radio 
interferometric technique, which provides a direct measure of 
the Fourier component of  
the sky brightness I($\bold{s}$) convolved with the primary beam of 
the antennas $B_{ij}(\bold{s})$ towards direction $\bold{s}$, often known
as the visibility $V_{ij}(\bold{u})$ at a given baseline $\bold{u}$ 
in units of wavelength:
\begin{equation}
V_{ij}(\bold{u})=g_ig_j^*\int B_{ij}(\bold{s})I(\bold{s})
                    e^{-2\pi i\bold{u}\cdot\bold{s}} d^2\bold{s},
\end{equation}
where $g_{i}$ and  $g_{j}$  are the complex gain factors of antenna pair 
$i$ and $j$, respectively. It can be easily shown that the angular power
spectrum ($C_{\ell}$) of the sky brightness distribution I($\bold{s}$) 
can be constructed by the average value of the square of $V_{ij}(\bold{u})$
with the Fourier wavenumber $\ell=2\pi u$ (White et al. 1999). 
In particular, under the assumption that the angular
power spectrum varies rather slowly with scale relative to 
the Fourier component of the primary beam 
$\tilde{B}_{ij}(\bold{u})={\cal F}[B_{ij}(\bold{s})]$,
we obtain the commonly used formula in the estimation of the angular power 
spectrum of low-frequency sky at a given frequency 
(Bharadwaj \& Sethi 2001; Zaldarriaga et al. 2004; Bharadwaj \& Ali 2005; 
Santos et al. 2005;  Ali et al. 2008; 
Pen et al. 2009; Paciga et al. 2011; Ghosh et al. 2011a,2011b):
\begin{equation}
\langle |V_{ij}(\bold{u})|^2\rangle\approx C_{\ell=2\pi u}\;
      |g_i|^2|g_j|^2 \int d^2\bold{u}'|\tilde{B}_{ij}(\bold{u}-\bold{u}')|^2.
\end{equation}
However, most of the theoretical studies in literature implicitly assume that 
perfect calibrations are being made for radio interferometers, and 
gains and primary beam introduce no spectral and spatial 
structures to destroy the reconstruction of the EoR angular power spectrum.  
But in reality, to perform precise calibration in the existence of 
ionospheric turbulence and to maintain high stability of radio instruments 
for rather a long integration time are extremely difficult for 
current 21cm experiments. This motivates us in this Letter 
to explore a possible remedy to overcome some of these shortcomings by 
using the visibility correlation coefficients instead of the visibilities. 
With the novel method it will be possible to statistically
extract the EoR angular power spectrum from radio interferometric 
measurements, independently of the primary beam of 
antennas. The method also allows us to remove partially the influence of 
receiver gains in the statistical measurement of $C_{\ell}$ because the
amplitudes of the gains cancel each other out.

\section {Formalism}

We begin with the auto-correlation of voltages   
measured at each antenna, which corresponds to the visibility 
with baseline $\bold{u}=0$:
\begin{equation}
|V_{ii}|^2= |g_i|^4 I_0^2 \int d^2\bold{s}|B_{ii}(\bold{s})|^2.
\end{equation}
This formula applies for discrete extragalactic sources only, and 
$I_0$ denotes the mean brightness of the point sources rather than
the total power.  The latter is dominated by the Milky Way in
our interested frequency range below 200 MHz. 
Now we define the visibility correlation coefficient such that 
\begin{equation}
p_{ij}(\bold{u})=\frac{V_{ij}(\bold{u})}{\sqrt{|V_{ii}||V_{jj}|}}.
\end{equation}
With this definition the amplitudes of the complex gains cancel 
each other out but the signal coherence remains.  Following eq.(2) 
we take the average value of the square of $p_{ij}(\bold{u})$
\begin{equation}
\langle |p_{ij}(\bold{u})|^2  \rangle
           \approx  \frac{C_{\ell=2\pi u}}{I_0^2}\;
            \frac{\int d^2\bold{u}'|\tilde{B}_{ij}(\bold{u}-\bold{u}')|^2}
                 {\sqrt{ \int d^2\bold{s}|B_{ii}(\bold{s})|^2\;
                         \int d^2\bold{s}|B_{jj}(\bold{s})|^2}}.
\end{equation}
Using the 'frequency shift' theorem of the Fourier transform, 
we can rewrite the Fourier component of the primary beam  
$\tilde{B}_{ij}(\bold{u}-\bold{u}')$ as 
$\tilde{B}_{ij}(\bold{u}-\bold{u}')={\cal F} [B_{ij}(-\bold{s})
         e^{-2\pi i\bold{u}\cdot\bold{s}}]$. 
According to the Parseval's theorem, the total power in the $\bold{s}$ 
domain or the $\bold{u}'$ domain should be the same. This yields
\begin{eqnarray}
\frac{1}{4\pi^2}\int  d^2\bold{u}' |\tilde{B}_{ij}(\bold{u}-\bold{u}')|^2 
             & =& \\ \nonumber
  \int  d^2\bold{s} |B_{ij}(-\bold{s})e^{-2\pi i\bold{u}\cdot\bold{s}}|^2 
     & = & \int  d^2\bold{s} |B_{ij}(-\bold{s})|^2. 
\end{eqnarray}
The most crucial point of this equation, however, is that the power 
$\int  d^2\bold{u}' |\tilde{B}_{ij}(\bold{u}-\bold{u}')|^2$ is 
actually independent of $\bold{u}$. Namely, in the conventional
estimate of angular power spectrum from eq.(2) the primary beam term 
$\int  d^2\bold{u}' |\tilde{B}_{ij}(\bold{u}-\bold{u}')|^2$ does not alter
the shape of the power spectrum. This arises, of course, from the presumption 
that the primary beam varies rather slowly with angular scale relative to the
cosmic signal, a necessary prerequisite for taking out from the integral 
the $C_{\ell=2\pi u}$ term in eq.(2).  
Furthermore, the symmetric feature of the primary beam for interferometric 
array element in all the current 21cm experiments suggests that 
$B_{ij}(\bold{s})$ can be actually treated as an even function,
implying $B_{ij}(-\bold{s})=B_{ij}(\bold{s})$.  Now, we apply the 
Cauthy-Schwarz inequality to the integral 
$\int  d^2\bold{s} |B_{ij}(\bold{s})|^2$ by noticing 
$B_{ij}(\bold{s})=B_{i}(\bold{s})B_{j}^*(\bold{s})$,  
$|B_{i}(\bold{s})|^2=|B_{ii}(\bold{s})|$ and 
$|B_{j}(\bold{s})|^2=|B_{jj}(\bold{s})|$,  and obtain 
\begin{eqnarray}
 \left|\int  d^2\bold{s} |B_{ij}(\bold{s})|^2 \right|^2
&= &\left|\int  d^2\bold{s} |B_{ii}(\bold{s})| |B_{jj}(\bold{s})| \right|^2
        \\ \nonumber
&\le &
\int  d^2\bold{s} |B_{ii}(\bold{s})|^2 \int  d^2\bold{s} |B_{jj}(\bold{s})|^2.
\end{eqnarray}
The equals sign holds when all antenna elements have 
identical primary beam $B_{i}(\bold{s})=B_{j}(\bold{s})$.  
Finally, replacing the integral 
$\int d^2\bold{u}'|\tilde{B}_{ij}(\bold{u}-\bold{u}')|^2$ in eq.(5) by 
eqs.(6) and (7) and using again the Parseval's theorem we find 
\begin{equation}
   \frac{\int d^2\bold{u}'|\tilde{B}_{ij}(\bold{u}-\bold{u}')|^2}
     {\sqrt{ \int d^2\bold{s}|B_{ii}(\bold{s})|^2\;
     \int d^2\bold{s}|B_{jj}(\bold{s})|^2}}=4\pi^2
\end{equation}
This allows us to evaluate the angular power spectrum $C_{\ell}$ 
relative to the mean brightness $I_0$ of point sources simply through
\begin{equation}
C_{\ell=2\pi u}=4\pi^2 I_0^2\; \langle|p_{ij}(\bold{u})|^2\rangle.
\end{equation}
Consequently, we may use the average sky brightness of extragalactic sources
over the primary beam 
of the antenna as the reference or calibration. While many 
sophisticated models have been constructed for low frequency 
sky for the purpose of foreground removals in 21cm 
experiments (e.g. de Oliveira-Costa et al. 2008; 
Jeli$\acute{c}$ et al. 2008; Wilman et al. 2008; Vernstrom et al. 2011; Kogut 2012),
for our purpose a simple power-law as the global sky brightness temperature 
model of extragalactic sources 
should suffice to calibrate the angular power spectrum at each frequency $\nu$: $T_0=13.7{\rm K}\left(\frac{\nu}{150{\rm MHz}} \right)^{-2.74}$.
This is the best-fit mean sky temperature over a field of view 
of $20^{\circ}\times20^{\circ}$ at the frequency range 100-200 MHz 
based on the numerical simulations by Wilman et al (2008) after  
bright sources with fluxes greater than 10 mJy at 150 MHz are removed
(see Section 3). Note that  $C_{\ell}$ will represent the angular power 
spectrum of the sky brightness temperature distribution  
if $I_0$ is replaced by
$T_0$ in eq.(9): $C_{\ell=2\pi u}=4\pi^2 T_0^2\langle|p_{ij}(\bold{u})|^2\rangle$.

\section {Simulation test}

In order to test the feasibility of extracting angular power
spectrum of low-frequency sky with the novel method, we use the 
simulated sky maps by Wilman et al. (2008) as the input and 
the 21 CentiMeter Array (21CMA) as the radio interferometer.  
Following the same algorithm of Zheng et al. (2012), we generate a set
of sky maps of $20^{\circ}\times20^{\circ}$ for different frequencies,
containing five distinct radio source types out to redshift of z=20. 
We estimate the flux of each radio source at different observing 
frequencies between 100 MHz and 200 MHz by a running 
power-law in frequency: $S\propto\nu^{\alpha+\Delta\alpha\lg(\nu/\nu_*)}$, 
in which $\nu_*$ is the  characteristic frequency and is taken to be 150
MHz in this study. Finally, we exclude all the bright sources 
with fluxes $S_{150{\rm MHz}}$ exceeding 10 mJy at 150 MHz which are 
assumed to be resolvable 
by current radio interferometers and can therefore be excised with 
existing algorithms.  We fix a $4096^2$ grid for each of the simulated 
image, which gives rise to an angular resolution of $0^{'}.3$ (see Figure 1).

21CMA, sited in Western China, is a ground-based meter-wave array designed
to probe the EoR operating at frequencies from 50 MHz to 200 MHz. 
The array consists of 80 pods (or stations) with 127 log-period antennas 
for each, which are deployed in two perpendicular arms 
along east-west and north-south directions, respectively. 
Spacings between pods are integral multiples of
20 m, with a maximum baseline of 2740 m along each baseline.  
The 21CMA redundancy is being used for the purposes of not only calibrations 
but also statistical measurement of the angular power spectrum of EoR 
at specific modes. In this work we choose the 40 pods of the east-west 
arm to generate the {\it uv} sampling towards the north celestial 
pole region, which reduces significantly the computing complexity in 
visibilities $V_{i,j}(u,v)$ since only two-dimensional Fourier transform 
is involved for such a configuration. 
The corresponding baseline (i.e. {\it uv} sampling) distribution of 
the 780 pod pairs as interferometers for the 21CMA east-west arm is 
shown in Figure 2, among which there are only 127 independent baselines.

The simulated sky map is convolved with the 21CMA primary beam 
$B_{i}(\bold{s})$ which is identical for all the pods and takes 
approximately a Gaussian function with 
FWHM$=4^{\circ}.26(\nu/100{\rm MHz})^{-1}$. 
We perform the Fourier transform of the
simulated sky map modulated by $B_{ij}(\bold{s})$ to produce the {\it uv} map
in terms of eq.(1) by setting $g_i=g_j=1$. The {\it uv} map
is further sampled by the 21CMA east-west 
baselines shown in Figure 2. We now calculate
the visibility correlation coefficients 
$p_{ij}(u,v)$ at each frequency channel 
for all the 780 baselines. This
yields a set of $p_{ij}(u,v)$ measures at 127 independent Fourier modes  
$\ell=2\pi\sqrt{u^2+v^2}$. 
Finally,  the angular power spectra at these specific modes can be obtained
using eq.(9), $C_{\ell=2\pi u}=4\pi^2 T_0^2\langle|p_{ij}(u,v)|^2\rangle$, 
in which the average 
sky brightness temperature for our case is determined by the unresolved
extragalactic sources $T_0=13.7{\rm K}(\nu/150{\rm MHz})^{-2.74}$  because
the Galactic foreground has not been included in the simulation and the
bright sources have been already removed.  

Figure 3 shows the angular power spectra of the simulated sky maps at four
frequencies ranging from 120 MHz to 180 MHz, together with the recovered
ones at specific modes sampled by the 21CMA east-west baselines. 
The former are constructed directly using the Fourier transform of the 
simulated sky images without inclusion of any instrumental effects, 
while the latter are the reconstructed results from the 21CMA 
'observations' based on the novel algorithm of eq.(9). 
It appears that the two results show a remarkably good agreement.  
We have also demonstrated the measurement errors assuming an integration
time of 300 days, an observing bandwidth of 1 MHz and the system noise 
of 300 K.
In particular, the noise level at $C_{\ell}$ has been suppressed by a factor of 
$1/\sqrt{N_{\ell}}$ for the redundant baseline of $N_{\ell}$ equally spaced 
pods. Large error bars at small- and large-$\ell$ ends  
 can be attributed to the arcminute-scale angular resolution 
due to the short baselines and the cosmic variance due to the small 
field-of-view of the 21CMA, respectively. 
For comparison the reconstructed angular power spectra directly from Figure 1 
are also plotted in Figure 3. It appears that 
the Gaussian beam alters only the amplitude rather than the shape of 
the angular power spectrum in terms of eqs.(2) and (6).  
As is shown in Figure 3, such an amplitude effect has been corrected 
for when the visibility correlation coefficients $p_{ij}(u,v)$ are used. 
Yet, in practice the primary beam can hardly be modeled by a perfect Gaussian 
function or other form of simple analytical function. A careful
calibration of the primary beam of antennas to a high degree of precision 
must be made. Imperfect and inaccurate calibrations of 
both spatial and spectral properties of the primary beam may lead to
significant errors in reconstruction of the power spectrum of 
EoR for 21cm experiments. 
Employment of the visibility correlation coefficients in the statistical 
study of EoR allows us to eliminate concern about the calibration
of the primary beam. 
While with our new algorithm we have successfully recovered the 
angular power spectra of the simulated radio foregrounds,  
the foregrounds should be eventually suppressed to the level 
below 10 mK, a minimum requirement for extracting statistically the 
signatures of EoR. This can be achieved, for example,   
using the foreground removal technique suggested recently by 
Cho et al.(2012), which works straightforwardly with the 
angular power spectrum. We have tested the technique and found that 
the foregrounds can indeed be subtracted to the level below 10 mK.

\section{Discussion and Conclusions}

Instead of directly employing the visibilities in conventional  
interferometric measurements of the statistical fluctuations of EoR 
suggested in literature,  we propose to work with the visibility 
correlation coefficients defined by 
$p_{ij}(u,v)=V_{ij}(u,v)/\sqrt{|V_{ii}||V_{jj}|}$. This allows us to
eliminate the effect of primary beams of antennas and also partially
reduce the influence of receiver gains on the statistical extraction
of the angular power spectrum of the low-frequency sky:
$C_{\ell=2\pi u}=4\pi^2 T_0^2\langle|p_{ij}(u,v)|^2\rangle$. Yet, we need to
calibrate the power spectrum using the average brightness temperature 
$T_0$ of extragalactic sources for low-frequencies below 
200 MHz. Observationally, $T_0$ has been determined so far to a 
degree of satisfaction at least for our purpose 
(e.g.,  Di Matteo et al. 2002). 

Introduction of the visibility correlation coefficient $p_{ij}(u,v)$ 
does not change the coherence of original signal. Hence the phase 
correction such as self-calibration and/or redundant calibration should be 
still made before combining $p_{ij}(u,v)$ data. 
Furthermore, bright sources in the field-of-view have to be removed 
to reduce the Poisson noise in computation of the angular power 
spectrum. This also implies that the sidelobes of bright sources still
remain as troublesome for reconstruction of angular power spectrum 
no matter whether $V_{ij}(u,v)$ or $p_{ij}(u,v)$ is used.  Another reason
that bright sources should be excised before recovery of the
angular power spectrum is the requirement of uniformality assumption 
for taking out from the integral the angular power spectrum term 
$C_{\ell}$ in eq.(2).   

We have tested the feasibility of the novel method using the simulated
sky maps of $20^{\circ}\times20^{\circ}$ for extragalactic sources 
in low-frequencies as the targets and 
21CMA as the radio interferometer.
We have successfully recovered the angular power spectra of the 
foregrounds at specific Fourier modes sampled by the 127 independent 
baselines of the 21CMA east-west arm, after the bright sources with fluxes
of $S_{150{\rm MHz}}\geq10$ mJy are removed.  While we have not 
included the gain fluctuations in sampling of the visibilities, 
the new method does allow us to remove the effect of the spatial response of 
the 21CMA antennas on the reconstruction of the angular power spectrum
of the low-frequency sky. 
In combination of various sophisticated foreground removal techniques  
developed in recent years especially in the power spectrum domain
(e.g. Cho et al. 2012), we should be able to subtract the foreground to  
the level for statistical detection of the 21cm signal from EoR.

\section{Acknowledgements}
We gratefully acknowledge the constructive suggestions and 
insightful comments by the referee. In particular we would like to 
thank Abhik Ghosh for kindly pointing out an error in Eq.(3)
in the first version of our manuscript.
This work was supported by the Ministry of Science and 
Technology of China, under grant No. 2009CB824900

\newpage

\begin{figure}
\plotone{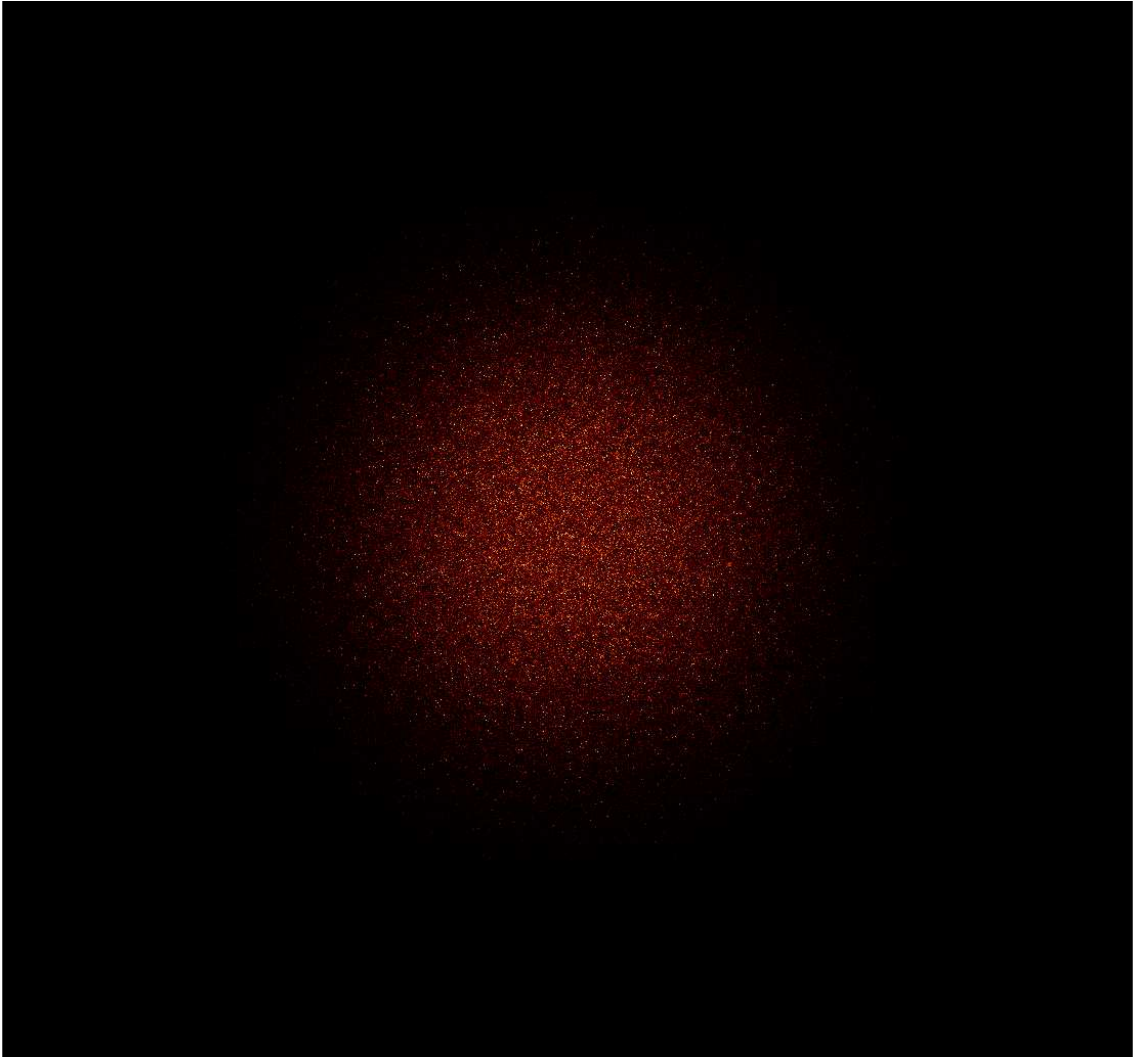}
\caption{Simulated sky map of $20^{\circ}\times20^{\circ}$
at $\nu=100$ MHz convolved with the 21CMA antenna primary beam. 
All the bright sources of $S_{150\rm{MHz}}\geq10$ mJy have been removed.}
\end{figure}

\begin{figure}
\plotone{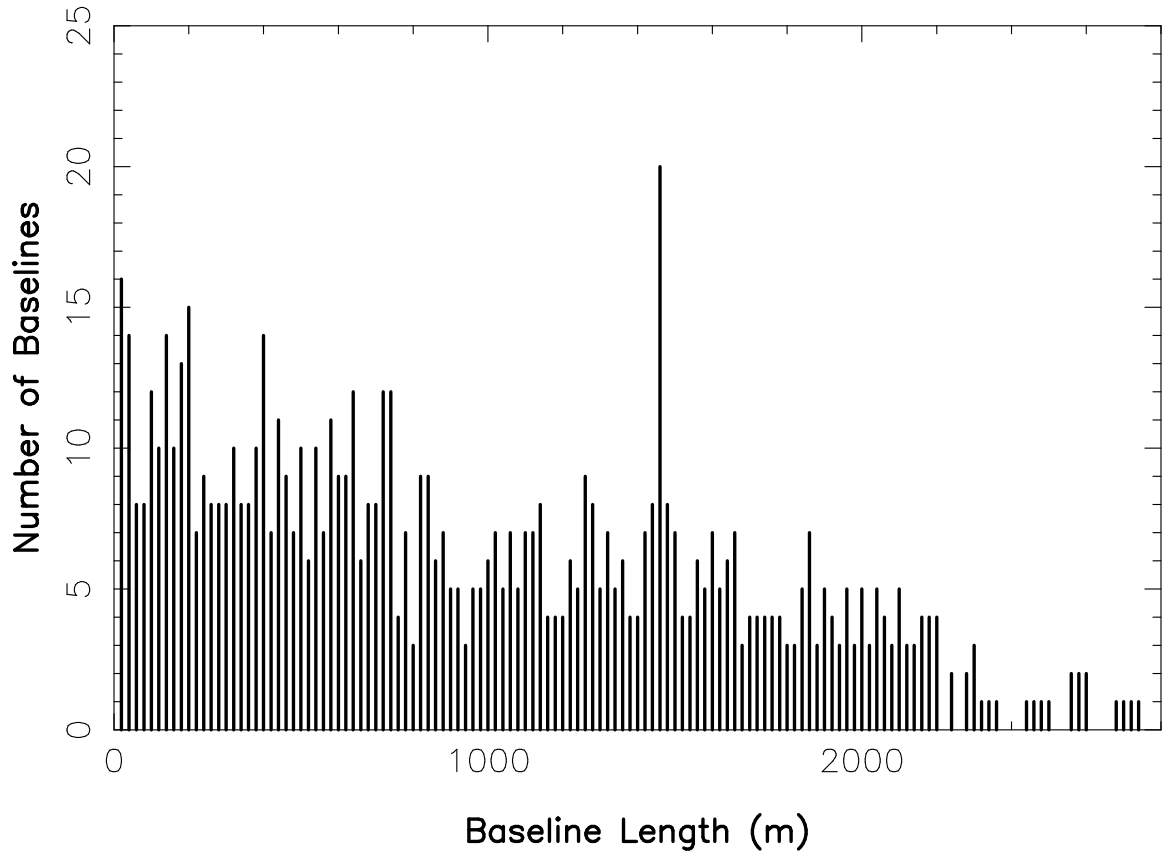}
\caption{Baseline distribution of the 780 pod pairs as interferometers  
in the 21CMA east-west arm, which also illustrates the {\it uv} sampling 
density when the horizontal axis is scaled in units of 
observing wavelength. Note that there are only 127 independent 
baselines among the 780 pairs.}
\end{figure}

\begin{figure}
\plotone{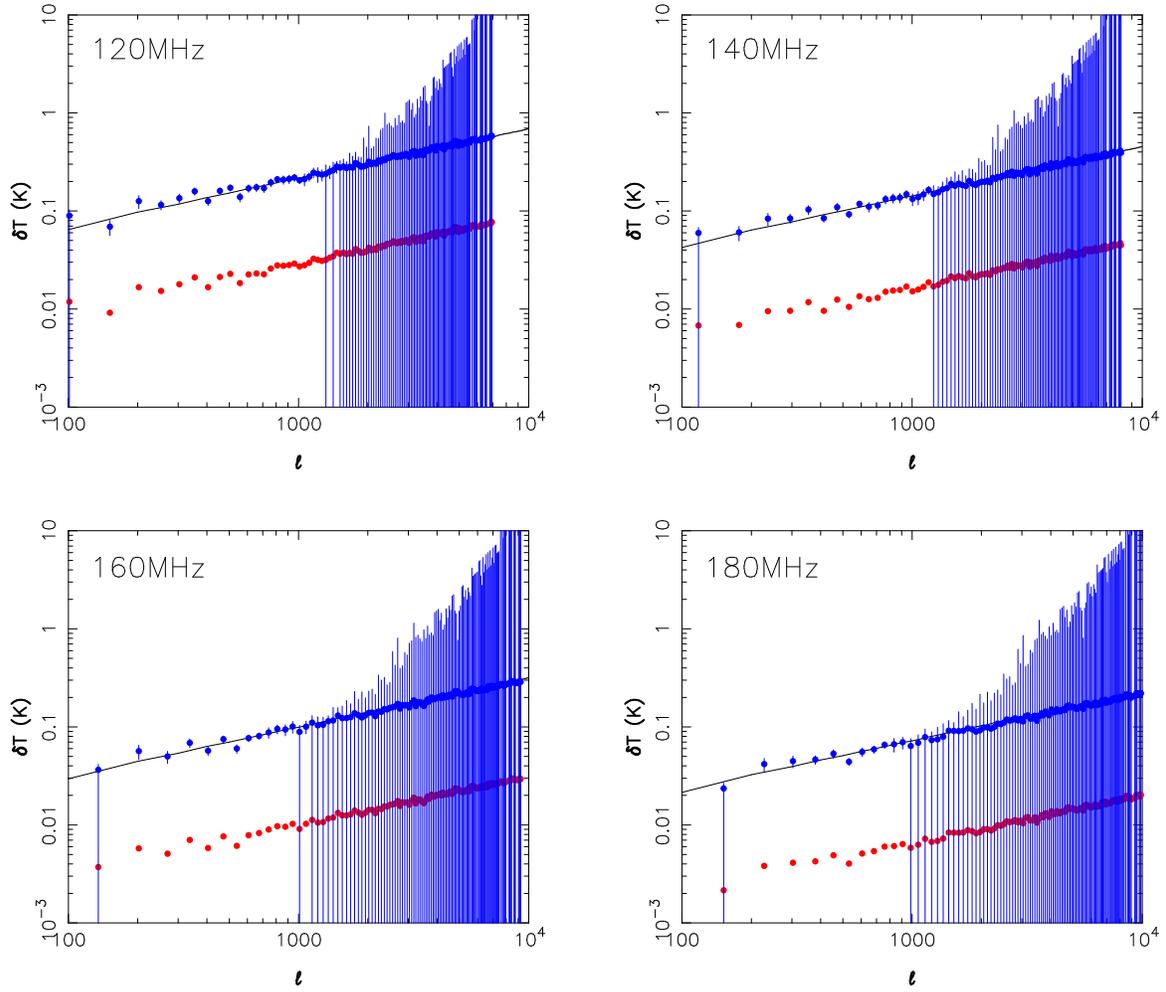}
\caption{Angular power spectra of the simulated sky maps at four 
frequencies ranging from 120 MHz to 180 MHz, represented by  
$\delta T=[\ell(2\ell+1)C_{\ell}/4\pi]^{1/2}$. 
Solid lines are the results derived from the simulated maps without
any observational and instrumental effects; blue circles are the 
reconstructed $\delta T$, in terms of the visibility correlation 
coefficients, measured at the 21CMA east-west independent 
baselines. For comparison 
the angular power spectra constructed directly from Fig.1 
are also shown (red circles), in which the Gaussian beam results in 
a decrease of the amplitude but does not alter the overall 
shape of $\delta T$. }
\end{figure}


\begin{references}
\reference{}Ali, S. S., Bharadwaj, S., \& Chengalur, J. N. 2008, \mnras, 
           385, 2166
\reference{}Bernardi, G., de Bruyn, A. G., Harker, G., et al. 2010, A\&A, 522, A67
\reference{}Bharadwaj, S., \& Ali, S. S. 2005, \mnras, 356, 1519
\reference{}Bharadwaj, S., \& Sethi, S. K. 2001, JA\&A, 22, 293
\reference{}Bowman, J. D., Morales, M. F.,  \& Hewitt, J. N. 2009, \apj, 695, 
            183
\reference{}Cho, J., Lazarian, A., \& Timbie, P. T.  2012, \apj, 749, 164
\reference{}Datta, A., Bhatnagar, S., \& Carilli, C. L. 2009, \apj, 703, 1851
\reference{}Datta, A., Bowman, J. D., \& Carilli, C. L. 2010, \apj, 724, 526
\reference{}de Oliveira-Costa, A., Tegmark, M., Gaensler, B. M., et al. 2008, MNRAS, 388, 247
\reference{}Di Matteo, T., Perna, R., Abel, T., \& Rees, M. J. 2002, \apj, 
            564, 576 
\reference{}Furlanetto, S. R., Oh, S. P., \& Briggs, F. H. 2006,
            Phys. Rep. 433, 181
\reference{}Ghosh, A., Bharadwaj, S., Ali, S. S., \& Chengalur, J. N. 2011a, 
            MNRAS, 411, 2426
\reference{}Ghosh, A., Bharadwaj, S., Ali, S. S., \& Chengalur, J. N. 2011b, 
            MNRAS, 418, 2584
\reference{}Jeli\'c V., Zaroubi, S., Labropoulos, P., et al. 2008, MNRAS, 389, 1319
\reference{}Kogut, A. 2012, \apj, 753, 110
\reference{}Liu, A.,  Tegmark, M., Bowman, J., Hewitt, J., \& Zaldarriaga, M. 
            2009, MNRAS, 398, 401 
\reference{}Madau, P., Meiksin, A., \& Rees, M. J. 1997, \apj, 475, 429 
\reference{}Morales, M. F., Bowmann, J. D., \& Hewitt, J. N. 2006, \apj, 
            648, 767 
\reference{}Morales, M. F., \& Wyithe, J. S. B. 2010, ARA\&A, 48, 127
\reference{}Paciga, G.,Chang, T. -C., Gupta, Y., et al. 2011, MNRAS, 413, 1174
\reference{}Pen, U. -L., Chang, T. -C., Hirata, C. M., et al. 2009, MNRAS, 399, 181
\reference{}Petrovic, N., \& Oh, S. P. 2011, MNRAS, 413, 2103
\reference{}Pritchard, J. R., \& Loeb, A. 2012, Rep. Prog. Phys., 75, 086901
\reference{}Santos, M. G., Cooray, A. \& Knox, L. 2005, ApJ, 625, 575
\reference{}Vernstrom, T., Scott, D., \& Wall, J. V. 2011, MNRAS, 415, 3641
\reference{}White, M., Carlstrom, J. E., Dragovan, M., \& Holzapfel, W. L. 1999, \apj, 514, 12
\reference{}Wilman, R. J., Miller, L., Jarvis, M. J., et al. 2008, MNRAS, 388, 1335
\reference{}Zaldarriaga, M., Furlanetto, S. R., \& Hernquist, L. 2004, 
\apj, 608, 622
\reference{}Zaroubi, S. 2012, arXiv:1206.0267
\reference{}Zheng, Q., Wu, X. -P., Gu, J. -H., Wang, J., \& Xu, H. 2012, 
            MNRAS, 424, 2562

\end{references}
\end{document}